\documentclass{article}

\usepackage{arxiv}

\usepackage[utf8]{inputenc}
\usepackage[T1]{fontenc}
\usepackage{hyperref}
\usepackage{url}
\usepackage{booktabs}
\usepackage{amsmath,amsfonts,amssymb}
\usepackage{graphicx}
\usepackage{xcolor}
\usepackage{tikz}
\usepackage{pgfplots}
\pgfplotsset{compat=1.18}
\usetikzlibrary{arrows.meta,positioning,calc,shapes.geometric}
\usepackage{placeins}
\usepackage{float}
\usepackage{listings}
\usepackage{caption}
\usepackage{dblfloatfix}
\usepackage[numbers,sort&compress]{natbib}
\usepackage{microtype}

\graphicspath{{figures/}}
\definecolor{singleShotColor}{RGB}{196,106,28}
\definecolor{pipelineColor}{RGB}{44,138,100}
\definecolor{linkBlue}{RGB}{28,67,125}

\hypersetup{%
    colorlinks=true,
    linkcolor=black,
    citecolor=black,
    urlcolor=linkBlue,
    pdfauthor={Chance A. LaVoie},
    pdftitle={Natural-Language to SysMLv2 Translation via Conformance-Driven Iterative Refinement},
    pdfkeywords={Systems Engineering, Design Automation, AI/KBS, Software Agents/Systems, Design Representation},
    pdfsubject={Systems Engineering}
}

\allowdisplaybreaks
\title{Natural-Language to SysMLv2 Translation via Conformance-Driven Iterative Refinement}

\author{%
    \begin{tabular}{c}
    \textbf{Chance LaVoie}$^{1}$ \quad
    \textbf{Eladio Andujar Lugo}$^{1}$\\[0.2em]
    \textbf{Taylan G. Topcu}$^{2}$ \quad
    \textbf{Levent Burak Kara}$^{1}$\\[0.55em]
    {\small $^{1}$Department of Mechanical Engineering, Carnegie Mellon University, Pittsburgh, PA 15213}\\
    {\small $^{2}$Grado Department of Industrial and Systems Engineering, Virginia Tech, Blacksburg, VA 24061}\\[0.25em]
    {\small Corresponding author: \texttt{lkara@andrew.cmu.edu}}
    \end{tabular}
}

\begin{document}
\maketitle
\begin{abstract}

Model-Based Systems Engineering (MBSE) relies on formal system models as primary technical artifacts for representing requirements, structure, and behavior across the system lifecycle. With the standardization of SysMLv2 as a textual language, interest is increasing in translating natural-language descriptions directly into executable models. For practical deployment, generated models must be accepted by industrial modeling environments, not merely satisfy grammar constraints. We present a conformance-checker-driven framework for reliable natural-language--to--SysMLv2 translation that enforces production-level acceptance as the termination condition. The system embeds a SysMLv2 conformance checker within a generate--check--repair loop. Each model is evaluated using the checker, and deterministic diagnostics are incorporated into revisions until zero conformance errors are achieved. Using the production checker as the oracle ensures the framework targets deployability rather than grammar plausibility. We evaluate the approach on the full SysMBench prompt set of 151 prompts across four large language model backends, yielding 604 prompt--model cases. Single-shot generation achieves 51.16\% production-conformance acceptance, while our approach achieves 100.00\% conformance. By elevating production conformance from a post-processing check to a control mechanism within generation, the framework converts probabilistic outputs into production-accepted SysMLv2 artifacts suitable for loading, visualization, and engineering use.

\end{abstract}

\keywords{Systems Engineering \and Design Automation \and AI/KBS \and Software Agents/Systems \and Design Representation}

\section{Introduction}

Model-Based Systems Engineering (MBSE) aims to replace document-centric workflows with model-centric engineering, where formal artifacts represent requirements, structure, behavior, and verification intent across the system lifecycle~\cite{estefan2008mbse,madni2018mbse}. By elevating the model to the primary technical authority, MBSE reduces cross-document reconciliation effort and improves traceability under change~\cite{madni2018mbse,incoseSEVision2035}. SysMLv2 strengthens this transition by standardizing a formal language with complementary textual and graphical representations and machine-readable artifacts under leadership of the Object Management Group~\cite{omgSysMLv2Spec2024}. In practice, engineers can author the same model textually and/or graphically, and exchange it across various engineering tools with reduced ambiguity. Because SysMLv2 is formal and textual, model construction is inherently scriptable. This scriptability creates an opportunity to translate engineering descriptions expressed in  natural-language directly into executable SysMLv2 artifacts. This is particularly promising for two main reasons. First, during the SysMLv1 era, MBSE adoption has been nascent by practitioners, primarily given perceptions of increased effort, time, and cost associated with SysML usage, along with lack of interoperability with other engineering tools \cite{henderson2021value, https://doi.org/10.1002/sys.21644}. Second, the widespread adoption of generative artificial intelligence capabilities, such as Large Language Models (LLMs) that natively operate in natural language, opens up new possibilities for efficiency \cite{mcdermott2020ai4se,husain_can_2024}. Hence, SysMLv2 has the potential to overcome these past obstacles. Nevertheless, for such translation to be useful in practice, generated models must be accepted by industrial modeling environments, not merely resemble well-formed text.

Recent advances in LLMs suggest that automated generation can meaningfully accelerate engineering workflows. In software engineering, controlled and field studies report substantial productivity gains from LLM-assisted generation, including 55.8\% faster task completion, 26.08\% higher weekly completed tasks in enterprise randomized deployments, and measurable increases in pull-request throughput~\cite{peng2023copilot,cui2025highSkilled,demirer2024fieldCopilot}. These results motivate analogous integration in MBSE, where automating low-value authoring steps could increase modeling speed and allow teams to focus on architectural and verification reasoning. 

\begin{center}
\captionsetup{type=figure,width=\textwidth}
\includegraphics[width=\textwidth,keepaspectratio]{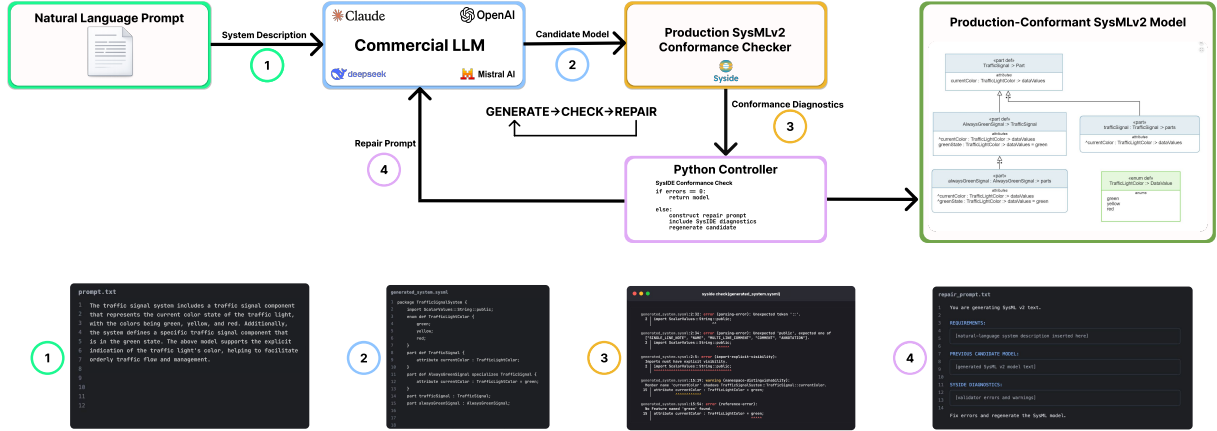}
\captionof{figure}{Conformance-checker-in-the-loop workflow for natural-language--to--SysMLv2 generation. A commercial LLM generates candidate SysMLv2 from a natural-language requirement, SysIDE evaluates the candidate for production conformance, and the resulting diagnostics are fed back to the model for iterative repair until a zero-error model is produced. Representative artifacts are shown for five stages of the loop: the input requirement, the generated SysMLv2 candidate model, the conformance diagnostics, the repair prompt, and the rendering of the final production-conformant SysMLv2 model.}
\label{fig:teaser_workflow}
\end{center}

\textbf{Current State of MBSE Automation:} Despite this opportunity, state-of-the-art LLMs do not reliably produce production-conformant\footnote{Production-conformant: accepted by the production SysMLv2 conformance checker with zero reported errors, indicating tool-level operational usability rather than grammar validity alone.} SysMLv2 models in single-shot generation. In response to this shortcoming, recent work has proposed structured pipelines for natural-language--to--SysMLv2 generation~\cite{bouamra2025systemp,cibrian2025sysmlagent}. SysTemp employs a template-first multi-agent design in which a template generator constructs a structured skeleton and a parser agent iteratively corrects syntax using grammar feedback~\cite{bouamra2025systemp}. Cibri{\'a}n et al.\ organize generation as an agentic loop with a retrieval-augmented context engine and ANTLR-based grammar validation as the primary syntax gate~\cite{cibrian2025sysmlagent}. In parallel, SysMBench establishes a benchmark for evaluating natural-language--to--SysML generation quality, especially with respect to semantic alignment~\cite{jin2025sysmbench}. Collectively, these efforts demonstrate that grammar-conforming SysMLv2 can be produced from natural language.

\textbf{Limits of Grammar-Level Parsing:} Despite this progress, grammar-level validity does not guarantee operational usability. Context-free grammar parsing ensures correct token ordering, balanced delimiters, and production-rule-compliant declarations. Industrial conformance checkers enforce additional model-wide constraints, including name resolution, type consistency, ownership rules, multiplicity constraints, and cross-reference integrity. A model that passes ANTLR parsing can still fail production conformance checks and therefore remain unusable for visualization, simulation, or downstream analysis. In an auxiliary repository demonstration, highlighted in Appendix A, we show ten distinct examples that pass SysML ANTLR parsing yet fail production conformance checks~\cite{LaVoieSupportingRepo}. The gap is therefore not grammar feasibility, but reliable generation of production-conformant SysMLv2 from arbitrary natural-language prompts.

\vspace{1\baselineskip}

\textbf{Digital Engineering and Trustworthy AI for Systems Engineering:} From a policy perspective, the proposed transition toward reliable natural-language--to--SysMLv2 model generation directly supports the strategic objectives in the Department of Defense (DoD) Digital Engineering Strategy ~\cite{office_of_the_deputy_assistant_secretary_of_defense_for_systems_engineering_dod_2018, office_of_the_under_secretary_of_defense_for_research_and_engineering_dod_2020}. DoD defines digital engineering as an integrated approach that uses authoritative system data and models across disciplines to support lifecycle activities from conceptualization through disposal ~\cite{office_of_the_deputy_assistant_secretary_of_defense_for_systems_engineering_dod_2018}. Central to this vision is establishing authoritative sources of truth (ASOT) that form the backbone of a digital thread linking every stage of the system lifecycle ~\cite{zimmerman_digital_2019}. As MBSE migrates to SysMLv2, which provides synchronized textual correspondence, SysMLv2 models are expected to form the core of ASOT. By enforcing production conformance as the termination criterion for generated artifacts, this work ensures AI-generated models meet the same conformance standards required for integration into digital engineering ecosystems. This supports long-term goals to shorten development timelines, improve design quality, and reduce total ownership cost ~\cite{topcu2025goldentriangle}.

However, integrating generative AI into systems engineering workflows raises questions of trust and reliability \cite{zhang2023trust}. Empirical studies show that LLMs exhibit common failure modes when generating systems engineering artifacts \cite{topcu_trust_2025}. These concerns are amplified because complex system developers, including government and industry, rely on legacy processes and domain experts embedded in established practices \cite{dane_reconsidering_2010}. Without mechanisms to verify AI-generated outputs against conformance criteria, engineering models risk being developed in isolation and integrated into digital ecosystems in ad hoc ways, increasing communication and synchronization burdens rather than reducing them \cite{robinson2012design}. Our conformance-driven approach addresses this trust barrier by automating production-level correctness. Instead of relying on probabilistic AI outputs and extensive review, the approach provides explicit conformance feedback at each iteration and requires zero-error conformance before handoff. This aligns with calls to measure digital engineering progress through concrete acceptance metrics ~\cite{henderson2023towards} and supports the broader research agenda on AI for systems engineering ~\cite{mcdermott2020ai4se}, where AI augmentation must be paired with verifiable correctness mechanisms for safe deployment and adoption.

\textbf{Proposed Advance:}
In this work, we present a conformance-checker--driven framework for natural-language--to--SysMLv2 generation that requires every generated model to pass a production conformance checker before it is considered complete. We place a production SysMLv2 conformance checker, SysIDE~\cite{sensmetry2024syside}, inside a generate--check--repair loop, shown in Figure~\ref{fig:teaser_workflow}. Given a natural-language prompt, the system produces a full textual SysMLv2 model and immediately checks it using the production conformance checker. The reported diagnostics, such as unresolved references, typing errors, and ownership violations, are then used to revise the model. We repeat this process until the conformance checker reports zero errors.

This approach follows the general idea of oracle-guided refinement from formal methods and inductive synthesis~\cite{clarke2000cegar,solarlezama2008thesis,jha2010ogis,alur2013sygus}, and is similar in spirit to compiler-feedback-driven code generation~\cite{wang2022compilable,grubisic2024compiler}. The difference here is that the oracle is a production modeling tool rather than a grammar parser. As a result, the loop enforces the same model-wide checks that an engineer would encounter when loading the model into an industrial environment.

We evaluate the framework on the full SysMBench prompt set of 151 prompts across four model backends, yielding 604 prompt--model cases. In single-shot generation, 51.16\% of the initial models are production-conformant. With conformance-guided refinement, all 604 cases reach zero-error conformance, with most resolving in only a few repair cycles. These results show that production conformance-checker feedback provides a reliable signal for correcting generated SysMLv2 models and makes direct integration into MBSE workflows feasible.
Our main contributions are:
\begin{enumerate}
    \item A conformance-checker-in-the-loop architecture for natural-language--to--SysMLv2 translation that enforces production conformance as a termination invariant.
    \item A benchmark-scale empirical study across the full SysMBench prompt set and four model backends that quantifies the gap between grammar validity and production conformance.
    \item A trajectory-level corpus of conformance-checker-guided refinement traces that enables future research on reliability, repair dynamics, and deployment-scale NL-to-MBSE systems.
\end{enumerate}
\section{Related Work}

Our review focuses on three areas that intersect in this study: (1) natural-language--to--SysML model generation within MBSE, (2) grammar-constrained and structured synthesis methods for large language models, and (3) verifier- and tool-guided refinement loops for generated artifacts. These bodies of work collectively demonstrate the feasibility of structured model synthesis from natural language and the benefits of deterministic feedback, while also highlighting open questions regarding deployment-level conformance in industrial modeling environments.

\subsection{Natural-Language to SysML Model Generation}

Recent research has explored the use of large language models to generate SysML and SysMLv2 artifacts directly from natural-language requirements. SysTemp~\cite{bouamra2025systemp} introduces a template-driven, multi-agent pipeline in which a structured model skeleton is first constructed and then refined through parser-guided correction. Cibri{\'a}n et al.~\cite{cibrian2025sysmlagent} propose an agent-based architecture that combines retrieval-augmented generation with ANTLR-based grammar validation to improve syntactic correctness. These systems demonstrate that structured prompting and iterative repair can substantially improve parser-level validity for SysMLv2 text.

Additional studies examine LLM behavior in SysML-related tasks more broadly. Wang et al.~\cite{wang2025sysmlbehaviorllmempirical} report empirical findings on the generation of SysML behavioral models and identify common inconsistencies and hallucinations in model structure. Other work investigates LLM-assisted interaction with SysMLv2 artifacts in engineering settings~\cite{dehart2024sysml_llm} and describes workflows in which generative AI is used to build and query MBSE models~\cite{arellano2015frameworks, extendingOntologyDrivenNL, fougeres2020intelligent, springer2025ontology}. These contributions reflect a growing interest in integrating LLMs into model-centric engineering practice.

Earlier efforts relied on restricted natural-language subsets, rule-based extraction, or diagram-level heuristics to translate requirements into SysML models~\cite{li2021restrictednlsysmlmodels,kbs2022nlpsystemsengineeringsysml,mistry2024texttomodeltransformationsysml}. While effective in constrained settings, these approaches often required controlled vocabularies or domain-specific grammars. More recent work explores LLM-assisted semantic alignment and model integration across SysMLv2 artifacts~\cite{anonymous2025semanticalignmentsysmlv2,hendricks2025texttomodel_sysml}. Collectively, this literature establishes the technical feasibility of NL-to-SysML pipelines, but most evaluations emphasize grammar conformity or diagram-level correctness rather than production-tool acceptance. Our work complements these efforts by examining acceptance under a production conformance-checker backend as the primary reliability criterion.

\subsection{Grammar-Constrained and Structured Synthesis}

Grammar-constrained decoding and template-based generation have emerged as practical strategies for improving structural correctness in LLM outputs. Grammar-constrained decoding methods restrict token sequences to context-free grammars, thereby enforcing well-formed structural patterns without requiring model retraining~\cite{geng2023gcd,park2024grammaraligneddecoding,frontiers2024gcdie}. These approaches reduce malformed outputs and improve syntactic validity across a range of structured generation tasks.

Template-driven methods further guide generation by providing predefined structural scaffolds or repair templates~\cite{aroOntologyNLP}. In SysML contexts, template-first pipelines such as SysTemp~\cite{bouamra2025systemp} use structured model outlines to stabilize generation. Such methods are particularly useful for newly standardized or low-data languages.

While grammar and template constraints effectively address token-level conformance, they do not necessarily enforce model-wide consistency or tool-specific constraints. Production modeling environments typically impose additional checks related to reference resolution, typing consistency, and cross-model integrity. As a result, grammar-level correctness represents an important but partial notion of conformance. Our approach builds on structured synthesis methods while shifting the acceptance criterion from parser conformity to production-tool conformance.

\subsection{Verifier- and Tool-Guided Refinement}

Iterative refinement using deterministic feedback has a long history in formal methods and program synthesis. Counterexample-guided abstraction refinement and oracle-guided synthesis frameworks use external checkers to iteratively improve candidate artifacts~\cite{clarke2000cegar,jha2010ogis,alur2013sygus}. In the context of large language models, Wang et al.~\cite{wang2022compilable} and Grubi{\v{s}}i{\'c} et al.~\cite{grubisic2024compiler} demonstrate that compiler diagnostics can serve as structured correction signals for generated code. Subsequent work extends this paradigm to test-driven and verifier-driven loops~\cite{ravi2025llmloop,sevenhuijsen2024vecogen,morvalho2025cegisprogramrepair,namin2024vermcts}.

Across these domains, a consistent pattern emerges: candidate artifacts are generated, evaluated by a deterministic backend, and revised until the backend reports success. In software engineering, the oracle is typically a compiler, test suite, or formal verifier. Our work instantiates this refinement pattern in the context of SysMLv2 MBSE, where the oracle is a production modeling tool. By aligning the termination condition with the same conformance mechanism used in industrial environments, we extend verifier-guided refinement to model-based systems engineering and examine its effect at benchmark scale.
\section{Methodology}

In this section, we describe the experimental design and the conformance-checker-in-the-loop procedure used to evaluate natural-language--to--SysMLv2 generation. We first define the paired study setup and then detail the iterative generation and conformance workflow.

\subsection{Study Objective and Paired Design}

Our goal is to evaluate whether placing a production conformance checker inside the generation loop improves reliability relative to single-shot generation. Specifically, we ask: for the same natural-language prompt and the same language model, does iterative conformance-guided refinement increase the rate of production-tool acceptance? In this study, we focus strictly on syntactic and production-level conformance.

The experimental unit is a unique prompt--model pair. For each pair, we compare two conditions obtained from the same generation trajectory:
\begin{enumerate}
    \item \textbf{Baseline (single-shot):} the production-conformance outcome of the initial model produced by the language model ($k=0$).
    \item \textbf{Pipeline (iterative):} the production-conformance outcome after applying conformance-guided repair until termination.
\end{enumerate}

Because both outcomes are derived from the same prompt and the same model instance, this paired design isolates the effect of iterative feedback. Prompt content and model identity remain fixed, and only the presence or absence of conformance-driven refinement differs between conditions.

\subsection{Conformance-checker-in-the-Loop Generation Procedure}

We implement a generate--check--repair workflow for natural-language--to--SysMLv2 translation. Given a prompt, the language model produces a complete textual SysMLv2 candidate. We then evaluate this candidate using a production conformance checker. If conformance errors are reported, we pass the deterministic diagnostics back to the model and request a revised candidate. This process continues until the checker reports zero errors.

For clarity, we formalize the repair loop in terms of the benchmark prompt, the checker output, and the repair prompt constructed at each cycle. Let $P$ denote the fixed natural-language system prompt from the SysMBench prompt set. At repair cycle $k$, the current SysMLv2 candidate $M_k$ is evaluated by the production conformance checker $C(\cdot)$, producing diagnostics
\[
D_k = C(M_k).
\]
The Python controller then deterministically constructs the repair prompt
\[
R_k = g\!\left(P, M_k, D_k\right),
\]
where $g(\cdot)$ denotes the controller's prompt-construction procedure. The language model generates the next candidate as
\[
M_{k+1} = f(R_k),
\]
where $f(\cdot)$ denotes the model generation call. In this formulation, $P$ is fixed across the full trajectory, while $M_k$, $D_k$, and $R_k$ are updated at each repair cycle. The diagnostics $D_k$ identify unresolved references, typing inconsistencies, ownership violations, and other model-wide constraint failures, providing the structured feedback used for the subsequent revision.

We use SysIDE (\texttt{syside check}) as the production conformance oracle because it exposes production SysMLv2 conformance checking through a scriptable command-line interface suitable for programmatic use inside the generate--check--repair loop~\cite{sensmetry2024syside}. A run terminates only when the checker reports zero errors. This choice is motivated by the distinction between grammar-level parsing and production conformance. In an auxiliary ten-case study included in our repository and appendix, we demonstrate that models can pass the SysMLv2 ANTLR4 (ANother Tool for Language Recognition, version 4) parser while still failing production conformance, illustrating that grammar conformity alone does not ensure operational usability~\cite{LaVoieSupportingRepo,hamrSysmlParser2026,sysmlV2PilotImplementation2026}. By using the production checker as the acceptance criterion, we align the loop with the same checks encountered in an industrial modeling environment. Figure~\ref{fig:teaser_workflow} summarizes the end-to-end conformance-checker-in-the-loop workflow and illustrates the representative artifacts exposed at each stage of generation, validation, and repair.

\subsection{Dataset and Outcome Extraction}
SysMBench provides paired natural-language prompts and ground-truth SysMLv2 models for benchmark evaluation~\cite{jin2025sysmbench}. In this study, we use the curated natural-language prompt set (IDs 1--151) as generation inputs because it was designed to stress SysMLv2 LLM generation across diverse modeling patterns.

To assess model-agnostic behavior of the same controller, we run four model configurations: OpenAI Codex 5.2 (\path|gpt-5.2-codex|)~\cite{openaiGPT52Codex2026}, Anthropic Sonnet 4.6 (\path|claude-sonnet-4-6|)~\cite{anthropicSonnet46API2026}, DeepSeek Reasoner v3.2 (\path|deepseek-reasoner|)~\cite{deepseekReasonerAPI2026}, and Mistral Large 2512 (\path|mistral-large-latest|)~\cite{mistralLargeLatestDocs2026}. This yields 604 prompt-level cases (151 prompts $\times$ 4 models).

\begin{figure}[H]
\centering
\definecolor{promptCardBg}{RGB}{248,247,243}
\definecolor{promptAccentOne}{RGB}{69,97,140}
\definecolor{promptAccentTwo}{RGB}{44,138,100}
\definecolor{promptAccentThree}{RGB}{196,106,28}
\definecolor{promptMetaText}{RGB}{92,99,108}
\begin{tikzpicture}
\tikzset{
    promptcard/.style={
        draw=black!16,
        fill=promptCardBg,
        rounded corners=6pt,
        line width=0.45pt,
        align=left,
        inner xsep=5pt,
        inner ysep=7pt,
        anchor=north west
    }
}

\node[promptcard] (p1) at (0,0) {%
    \begin{minipage}[t][48mm][t]{0.282\textwidth}
    \raggedright
    {\scriptsize\bfseries\textcolor{promptAccentOne}{Compact workflow}}\hfill
    {\tiny\color{promptMetaText}Prompt 33}\par
    \vspace{1pt}
    {\scriptsize\itshape\color{promptMetaText}Photography Technique}\par
    \vspace{3pt}
    {\fontsize{6.45}{7.7}\selectfont
    The system is designed to implement a camera information processing workflow. When a user selects a scene through the camera's viewfinder (viewPort), the system first focuses on the scene to obtain an image (Image). This image is then captured to generate a photograph (Picture). After the photograph is generated, the system displays it on the screen via the display port (displayPort).\par}
    \end{minipage}
};

\node[promptcard, right=2.0mm of p1] (p2) {%
    \begin{minipage}[t][48mm][t]{0.282\textwidth}
    \raggedright
    {\scriptsize\bfseries\textcolor{promptAccentTwo}{Spatiotemporal simulation}}\hfill
    {\tiny\color{promptMetaText}Prompt 114}\par
    \vspace{1pt}
    {\scriptsize\itshape\color{promptMetaText}Vehicle Traffic}\par
    \vspace{3pt}
    {\fontsize{6.45}{7.7}\selectfont
    This system is designed for spatiotemporal simulation of the dynamic behavior of vehicles on roads at different moments. Users can define parameters such as the vehicle's mass, position, velocity, and acceleration, and, combined with the road's slope (angle) and surface friction coefficient, depict the state of the vehicle and the road at specific time points. The system supports snapshot recording at multiple moments within the simulation time series, enabling tracking of the vehicle's state transitions from start-up (on state), through the driving process, to shutdown (off state).\par}
    \end{minipage}
};

\node[promptcard, right=2.0mm of p2] (p3) {%
    \begin{minipage}[t][48mm][t]{0.282\textwidth}
    \raggedright
    {\scriptsize\bfseries\textcolor{promptAccentThree}{Safety-critical monitoring}}\hfill
    {\tiny\color{promptMetaText}Prompt 134}\par
    \vspace{1pt}
    {\scriptsize\itshape\color{promptMetaText}Medical Health}\par
    \vspace{3pt}
    {\fontsize{6.45}{7.7}\selectfont
    This system is designed to ensure high reliability and safety of the blood glucose meter during use. When the battery is depleted or cannot be charged, the system should be able to automatically detect the battery status and promptly alert the user to prevent failure to measure blood glucose levels due to battery issues, as well as potential treatment delays resulting from such failures. To prevent the aforementioned failure scenarios, the system requires the implementation of preventive measures for battery status, and it must have appropriate alarm and emergency response mechanisms in case of abnormalities in the blood glucose measurement function.\par}
    \end{minipage}
};

\draw[promptAccentOne, line width=1.7pt] ([xshift=5pt,yshift=-5pt]p1.north west) -- ([xshift=-5pt,yshift=-5pt]p1.north east);
\draw[promptAccentTwo, line width=1.7pt] ([xshift=5pt,yshift=-5pt]p2.north west) -- ([xshift=-5pt,yshift=-5pt]p2.north east);
\draw[promptAccentThree, line width=1.7pt] ([xshift=5pt,yshift=-5pt]p3.north west) -- ([xshift=-5pt,yshift=-5pt]p3.north east);
\end{tikzpicture}
\caption{Representative SysMBench prompt excerpts illustrating the variety and structure of the benchmark inputs. The examples show a compact workflow prompt (Prompt 33), a spatiotemporal vehicle-simulation prompt (Prompt 114), and a safety-critical monitoring prompt (Prompt 134).}
\label{fig:prompt_showcase}
\end{figure}
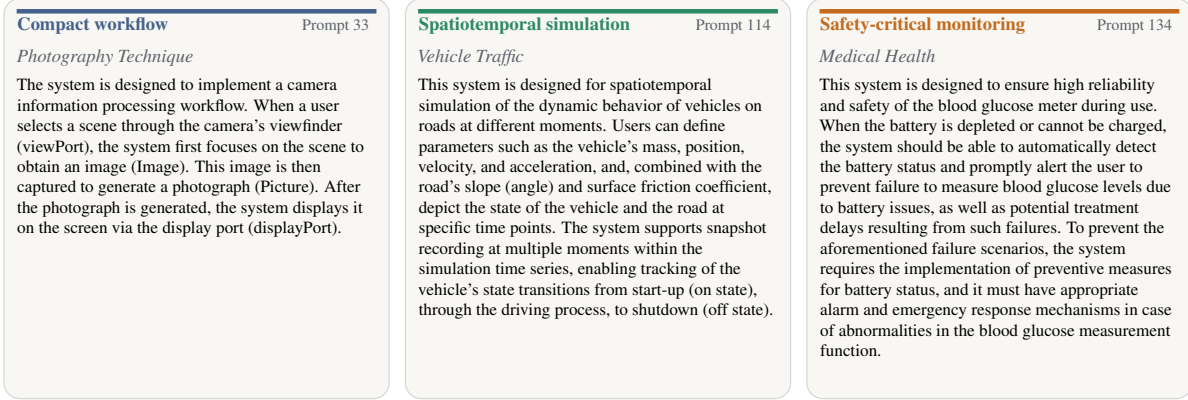

From saved run records, we extract single-shot and pipeline pass/fail outcomes, iterations run, iterations to success, first/final error counts, cumulative error counts, per-iteration runtime, and token usage. For grammar-versus-production analysis, we also extract single-shot ANTLR parse pass/fail and single-shot SysIDE pass/fail from the same initial candidate artifacts for all 604 prompt--model cases. Representative examples of these benchmark prompts are shown in Figure~\ref{fig:prompt_showcase}.

\subsection{Evaluation Metrics}
\subsubsection{Production Conformance}
The primary evaluation metrics are production conformance rates. For each analysis slice (overall and per-model), we report:
\begin{enumerate}
    \item single-shot conformance rate (initial candidate, $k=0$),
    \item pipeline conformance rate (final iteration of the conformance-gated loop).
\end{enumerate}
We also report iterations-to-success, defined as the number of repair cycles required for a prompt--model case to reach production conformance.

\subsubsection{Convergence Metrics}
To characterize how quickly the loop approaches acceptance, we index progress by repair cycles. Let $k=0$ denote the initial single-shot generation (no checker feedback), and let $k\geq 1$ denote $k$ rounds of generate--check--repair. Let the $N$ prompt--model cases be indexed by $i=1,\ldots,N$, and let $s_i$ denote the first repair cycle at which case $i$ reaches production conformance.
 We define
\[
A_k = \frac{1}{N}\sum_{i=1}^{N}\mathbf{1}[s_i \le k],
\]
the cumulative fraction of prompt--model cases that have reached production conformance by repair cycle $k$.

We define residual failure mass as
\[
R_k = 1 - A_k,
\]
where $R_k$ is the fraction of prompt--model cases not yet accepted at repair cycle $k$. We then define an empirical contraction ratio
\[
\rho_k = \frac{R_{k+1}}{R_k}.
\]
When $0 < \rho_k < 1$, residual failure shrinks from one cycle to the next; when $\rho_k$ is approximately stable across $k$, the observed trajectory suggests contraction-like behavior, with approximately multiplicative reduction in the early repair regime under standard contraction-style analyses in iterative methods~\cite{banach1922operations,nocedal2006numerical,he2016averageConvergenceRate}. In finite empirical campaigns, we estimate contraction behavior from early-to-mid cycles, before residual mass becomes very small; terminal transitions are excluded from rate estimation because finite-sample tail effects can produce unstable ratios near the finite-sample resolution. This characterization is descriptive of observed finite-sample behavior and does not assert formal convergence guarantees.

We also report time-to-threshold acceptance
\[
T_{\varepsilon} = \min\{k : A_k \ge 1-\varepsilon\},
\]
with explicit reporting of $T_{90}$, $T_{95}$, and $T_{99}$. This follows standard iteration-to-threshold (iteration-to-$\varepsilon$) complexity summaries in optimization~\cite{polyak1987introductionOptimization}. These metrics provide an interpretable rate summary complementary to pass-rate metrics.

This framing follows empirical convergence-rate characterizations used in iterative optimization and search, including multiplicative (geometric-style) interpretations of residual reduction~\cite{he2016averageConvergenceRate}, while remaining descriptive rather than proving formal convergence guarantees. It is also consistent with iterative verifier-feedback refinement in code generation, where deterministic diagnostics guide successive corrections toward acceptance~\cite{wang2022compilable,grubisic2024compiler}.

\subsubsection{Statistical Reliability Analysis}
For convergence reliability, let $Y_i$ indicate whether prompt--model case $i$ reaches production conformance within the allowed repair loop:
\[
Y_i =
\begin{cases}
1, & \text{case } i \text{ reaches zero SysIDE errors},\\
0, & \text{otherwise}.
\end{cases}
\]
We model $Y_i$ as a Bernoulli outcome with convergence probability $p$ under the evaluated controller, conformance checker, and model backends. Because the loop stops once a candidate passes the checker, all initially conformant cases remain final successes. With success observed for all $n$ evaluated prompt--model cases, we report the exact one-sided 95\% Clopper--Pearson lower confidence bound, which is appropriate for the all-success case:

\[
p_L = \alpha^{1/n}, \qquad \alpha = 0.05.
\]
Equivalently, for failure probability $q=1-p$, the corresponding one-sided upper bound is
\[
q_U = 1-p_L = 1-\alpha^{1/n}
      \approx \frac{-\ln(\alpha)}{n}.
\]
These bounds are scoped to SysMBench-style prompt distributions under the evaluated configuration and are reported numerically in Section~\ref{sec:results}. They are not treated as universal guarantees over arbitrary natural-language inputs.

\subsubsection{Grammar vs. Production Conformance}
To quantify the operational gap between grammar-level validity and production conformance, we compare ANTLR and SysIDE outcomes on the same single-shot artifact from each case. This comparison is performed on the initial candidate ($k=0$ in repair-cycle indexing).

Grammar validity is defined as ANTLR parse success using the SysMLv2 ANTLR4 parser, which provides a generated grammar and Java parser/lexer derived from the SysML v2 Pilot Implementation~\cite{hamrSysmlParser2026,sysmlV2PilotImplementation2026}. This ANTLR check is used only as a representative context-free grammar parsability test and does not enforce production model-wide constraints. Production conformance is defined as SysIDE acceptance (zero SysIDE errors) under \texttt{syside check}. For each case, we record binary pass/fail outcomes for both checks and construct a $2\times2$ contingency table (ANTLR pass/fail $\times$ SysIDE pass/fail).

\subsection{Extended Analyses}
In addition to the primary evaluation metrics described above, we conduct several supplementary analyses to better understand repair behavior within the conformance-guided generation loop. These analyses are presented in Appendix~B and Appendix~C. Specifically, we examine relationships between iterations-to-success and generated output length as well as SysMBench benchmark difficulty labels (Appendix~B). We also analyze persistent conformance errors across repair iterations to characterize cases in which the language model fails to resolve a reported error despite attempting repair (Appendix~C). These analyses provide additional insight into repair dynamics and factors influencing convergence behavior.

\section{Results}
\label{sec:results}

In this section, we present the results of our evaluation of production conformance outcomes across the benchmark. We begin by examining overall acceptance behavior and then analyze reliability, convergence dynamics, and conformance characteristics across models.

\subsection{Primary Outcome: Production Conformance}
\begin{figure}[H]
\centering
\definecolor{singleShotColor}{RGB}{196,106,28}
\definecolor{pipelineColor}{RGB}{44,138,100}
\begin{tikzpicture}
\begin{axis}[
    xbar,
    width=0.64\columnwidth,
    height=0.18\columnwidth,
    xmin=0,
    xmax=105,
    xlabel={Acceptance Rate (\%)},
    symbolic y coords={Pipeline,Single-shot},
    ytick={Single-shot,Pipeline},
    yticklabels={Single-shot,Pipeline},
    enlarge y limits={abs=4pt},
    axis lines*=left,
    xmajorgrids=true,
    grid style={dashed,gray!25},
    tick label style={font=\footnotesize},
    label style={font=\footnotesize},
    nodes near coords,
    every node near coord/.append style={font=\footnotesize, text=black, anchor=west, xshift=2pt},
]
\addplot+[fill=singleShotColor, draw=singleShotColor!70!black] coordinates {
    (51.16,Single-shot)
};
\addplot+[fill=pipelineColor, draw=pipelineColor!70!black] coordinates {
    (100.00,Pipeline)
};
\end{axis}
\end{tikzpicture}
\caption{Overall single-shot vs. final pipeline production conformance across all 604 prompt-level cases.}
\label{fig:overall_baseline_vs_pipeline}
\end{figure}
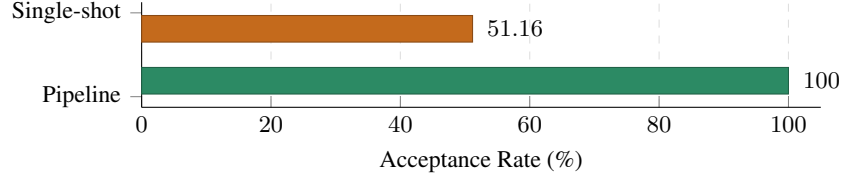

Across all 604 prompt-level trials (151 prompts for each of four models), single-shot outputs conformed for 309/604 cases (51.16\%). With the conformance-checker-in-the-loop, the pipeline outputs conformed for 604/604 cases (100.00\%). Figure~\ref{fig:overall_baseline_vs_pipeline} summarizes this overall single-shot versus final pipeline comparison.

\FloatBarrier

\subsection{Per-Model Reliability}
\begin{figure}[H]
\centering
\begin{tikzpicture}
\begin{axis}[
    ybar,
    bar width=7pt,
    width=0.58\columnwidth,
    height=0.30\columnwidth,
    ymin=0,
    ymax=110,
    ylabel={Acceptance Rate (\%)},
    symbolic x coords={Sonnet,OpenAI,DeepSeek,Mistral},
    xtick=data,
    xticklabel style={font=\footnotesize, rotate=0, anchor=north},
    nodes near coords,
    nodes near coords={\pgfmathprintnumber[fixed,precision=2]{\pgfplotspointmeta}},
    nodes near coords align={center},
    nodes near coords style={
        font=\tiny,
        text=black,
    },
    every node near coord/.append style={anchor=south, yshift=-1.8pt},
    ymajorgrids=false,
    enlarge x limits=0.16,
    axis lines*=left,
    tick label style={font=\footnotesize},
    label style={font=\footnotesize},
    legend style={
        draw=none,
        font=\footnotesize,
        at={(0.5,1.04)},
        anchor=south,
        legend columns=2
    },
    legend cell align={left},
    legend image code/.code={
        \draw[#1] (0cm,-0.075cm) rectangle (0.22cm,0.075cm);
    },
]
\addplot+[fill=singleShotColor, draw=singleShotColor!70!black] coordinates {
    (Sonnet,82.78)
    (OpenAI,41.72)
    (DeepSeek,41.06)
    (Mistral,39.07)
};
\addplot+[fill=pipelineColor, draw=pipelineColor!70!black] coordinates {
    (Sonnet,100.00)
    (OpenAI,100.00)
    (DeepSeek,100.00)
    (Mistral,100.00)
};
\legend{Single-shot ($k=0$),Pipeline (final)}
\end{axis}
\end{tikzpicture}
\caption{Per-model single-shot vs. final pipeline production conformance.}
\label{fig:by_model_baseline_vs_pipeline}
\end{figure}
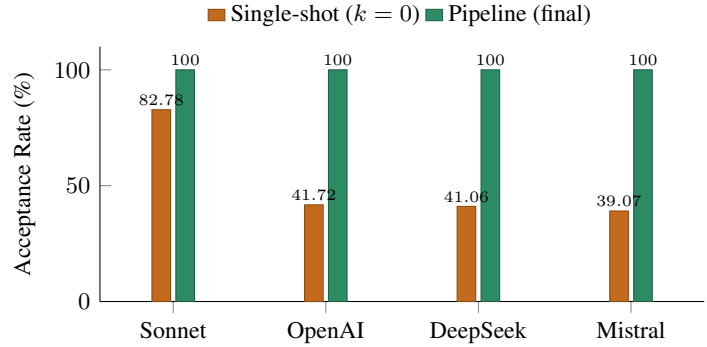

All models reached 151/151 eventual production conformance under the conformance-gated loop, but single-shot pass rates differed substantially. Figure~\ref{fig:by_model_baseline_vs_pipeline} provides the per-model comparison with exact conformance values annotated on the bars.

Anthropic Sonnet 4.6 had the highest single-shot conformance rate (82.78\%), while OpenAI Codex 5.2 (41.72\%), DeepSeek Reasoner (41.06\%), and Mistral Large (39.07\%) showed worse single-shot behavior and larger single-shot-to-pipeline gaps.

\FloatBarrier

\subsection{Convergence Behavior}
Convergence is indexed by repair cycles: $k=0$ denotes the initial single-shot generation (no conformance feedback), and $k\geq 1$ denotes conformance-guided repair cycles; acceptance corresponds to zero-error output under the production checker.

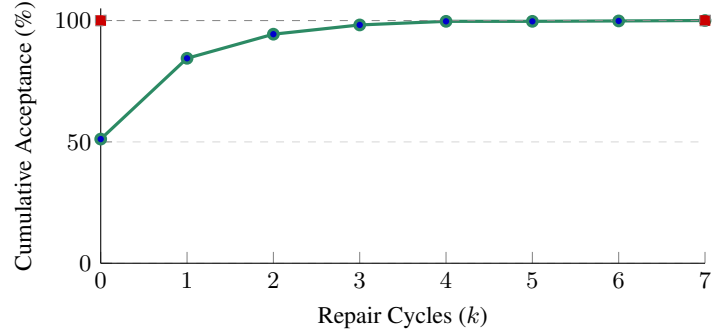
\begin{figure}[H]
\centering
\begin{tikzpicture}
\begin{axis}[
    width=0.58\columnwidth,
    height=0.30\columnwidth,
    xmin=0,
    xmax=7,
    ymin=0,
    ymax=105,
    xlabel={Repair Cycles ($k$)},
    ylabel={Cumulative Acceptance (\%)},
    xtick={0,1,2,3,4,5,6,7},
    ymajorgrids=true,
    xmajorgrids=false,
    grid style={dashed,gray!30},
    axis lines*=left,
    tick label style={font=\footnotesize},
    label style={font=\footnotesize},
]
\addplot+[pipelineColor, very thick, mark=*, mark size=1.8pt] coordinates {
    (0,51.16)
    (1,84.44)
    (2,94.37)
    (3,98.18)
    (4,99.67)
    (5,99.67)
    (6,99.83)
    (7,100.00)
};
\addplot+[black!45, dashed, thin] coordinates {(0,100) (7,100)};
\end{axis}
\end{tikzpicture}
\caption{Cumulative production conformance versus repair cycles ($k$). $k=0$ denotes initial single-shot generation; $k\geq 1$ denotes conformance-guided repair cycles. Conformance corresponds to zero-error output under the production checker.}
\label{fig:cumulative_convergence}
\end{figure}

Figure~\ref{fig:cumulative_convergence} first presents the pooled convergence trajectory across all prompt-level cases.

The pooled curve shows a large first-step jump from $k=0$ to $k=1$, followed by rapid compression by $k=2$ and then a short tail. Table~\ref{tab:convergence_behavior} provides the exact counts and percentages underlying Figure~\ref{fig:cumulative_convergence}: acceptance increases from 51.16\% at $k=0$ to 84.44\% at $k=1$, reaches 94.37\% by $k=2$, and reaches 99.67\% by $k=4$. Only two cases remain in the grouped $k=5$--$7$ tail (0.33\%), where cumulative acceptance reaches 100.00\%. Consistent with this front-loaded pattern, total attempts to first conformance, counting the initial single-shot attempt, are summarized by mean 1.727, median 1, IQR 1--2, and maximum 8; equivalently, repair cycles to first conformance are summarized by mean 0.727, median 0, IQR 0--1, and maximum 7.

\begin{table}[H]
\centering
\caption{Distribution of repair cycles to first production conformance (pooled across 604 prompt-level cases).}
\label{tab:convergence_behavior}
\begin{tabular}{lrrr}
\toprule
Repair cycles ($k$) & Cases & Share & Cumulative \\
\midrule
0 & 309 & 51.16\% & 51.16\% \\
1 & 201 & 33.28\% & 84.44\% \\
2 & 60 & 9.93\% & 94.37\% \\
3 & 23 & 3.81\% & 98.18\% \\
4 & 9 & 1.49\% & 99.67\% \\
5--7 & 2 & 0.33\% & 100.00\% \\
\bottomrule
\end{tabular}
\end{table}

In Table~\ref{tab:convergence_behavior}, Share denotes the fraction of cases first accepted at cycle $k$, while Cumulative denotes the total fraction accepted by cycle $k$.

\noindent\textbf{Rate Characterization.} To quantify the observed convergence shape, we report an empirical contraction analysis of residual failure mass. Under contraction-style iteration analysis, let residual failure mass be $R_k = 1 - A_k$. The observed residual sequence is $R_0=0.4884$, $R_1=0.1556$, $R_2=0.0563$, $R_3=0.0182$, and $R_4=0.0033$. Using early cycles, the empirical contraction ratios are $\rho_0 \approx 0.1556/0.4884 \approx 0.32$, $\rho_1 \approx 0.0563/0.1556 \approx 0.36$, and $\rho_2 \approx 0.0182/0.0563 \approx 0.32$. These values cluster around an average early-cycle contraction factor of approximately 0.33 (computed as the geometric mean of $\rho_0$--$\rho_2$). Tail transitions are excluded from contraction-rate estimation because ratios become unstable when residual mass is near the finite-sample resolution. These early-cycle ratios cluster in a narrow band, suggesting an approximately multiplicative ("contraction-like") reduction pattern in the initial repair regime. In practical terms, early cycles remove roughly two-thirds of the remaining failures per cycle in the observed early regime.

Complementing the contraction estimate, time-to-threshold metrics quantify convergence speed. The observed thresholds are $T_{90}=2$, $T_{95}=3$, and $T_{99}=4$. These values characterize how rapidly conformance-guided refinement approaches the acceptance set in repair-cycle space.

\begin{figure}[H]
\centering
\begin{tikzpicture}
\definecolor{sonnetColor}{RGB}{31,119,180}
\definecolor{openaiColor}{RGB}{255,127,14}
\definecolor{deepseekColor}{RGB}{44,160,44}
\definecolor{mistralColor}{RGB}{214,39,40}
\begin{axis}[
    width=0.60\columnwidth,
    height=0.32\columnwidth,
    xmin=0,
    xmax=7,
    ymin=0,
    ymax=102,
    xlabel={Repair Cycles ($k$)},
    ylabel={Cumulative Acceptance (\%)},
    xtick={0,1,2,3,4,5,6,7},
    ymajorgrids=true,
    xmajorgrids=false,
    grid style={dashed,gray!30},
    axis lines*=left,
    tick label style={font=\footnotesize},
    label style={font=\footnotesize},
    legend style={
        draw=none,
        font=\footnotesize,
        at={(0.5,1.02)},
        anchor=south,
        legend columns=2
    },
]
\addplot+[sonnetColor, very thick, mark=*] coordinates {
    (0,82.78) (1,97.35) (2,99.34) (3,99.34) (4,99.34) (5,99.34) (6,100.00) (7,100.00)
};
\addplot+[openaiColor, very thick, mark=square*] coordinates {
    (0,41.72) (1,88.08) (2,98.01) (3,99.34) (4,100.00) (5,100.00) (6,100.00) (7,100.00)
};
\addplot+[deepseekColor, very thick, mark=triangle*] coordinates {
    (0,41.06) (1,76.82) (2,89.40) (3,97.35) (4,99.34) (5,99.34) (6,99.34) (7,100.00)
};
\addplot+[mistralColor, very thick, mark=diamond*] coordinates {
    (0,39.07) (1,75.50) (2,90.73) (3,96.69) (4,100.00) (5,100.00) (6,100.00) (7,100.00)
};
\legend{Anthropic Sonnet 4.6,OpenAI Codex 5.2,DeepSeek Reasoner,Mistral Large}
\end{axis}
\end{tikzpicture}
\caption{Per-model cumulative production conformance versus repair cycles ($k$).}
\label{fig:cumulative_convergence_by_model}
\end{figure}
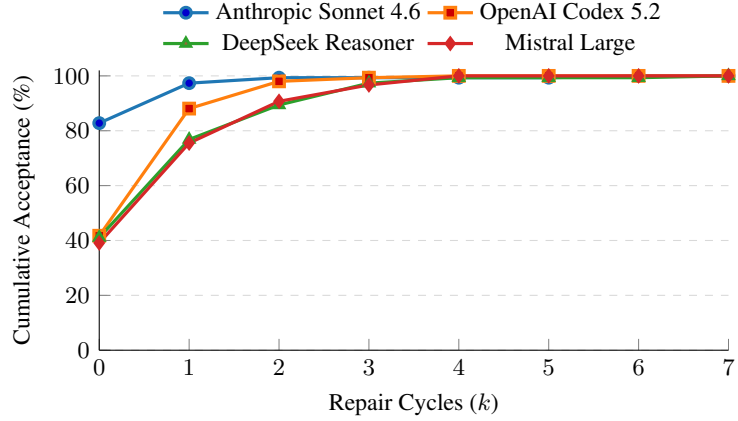

To assess whether this pooled behavior is shared across backends, Figure~\ref{fig:cumulative_convergence_by_model} overlays per-model cumulative acceptance trajectories.

Figure~\ref{fig:cumulative_convergence_by_model} shows that single-shot starting points differ across models, but the trajectories compress rapidly once conformance-guided repair begins. Despite different initial conformance levels, all curves move quickly toward full conformance within a small number of repair cycles. This pattern supports the model-agnostic control-signal interpretation: deterministic conformance diagnostics drive the dominant convergence dynamics across backends.

\FloatBarrier

\subsection{Statistical Reliability}

We now quantify the reliability of conformance-guided convergence across the full campaign. In total, we evaluated $n=604$ prompt--model cases. In single-shot generation, 309 of 604 cases (51.16\%) conformed. Under conformance-guided refinement, all 604 cases ultimately reached zero-error production conformance, with no observed failures.

To interpret this result statistically, we model convergence as a Bernoulli process and compute an exact one-sided 95\% Clopper--Pearson lower confidence bound on the convergence probability $p$~\cite{clopper1934confidenceLimits}. For the all-success case ($x=604$ successes in $n=604$ trials), the lower bound is
\[
p_L = \alpha^{1/n},
\]
with $\alpha=0.05$. Substituting $n=604$ gives
\[
p_L = (0.05)^{1/604} \approx 0.9951.
\]
With 95\% confidence, the true convergence probability for prompt--model cases drawn from this evaluated four-backend mixture is therefore at least 99.51\%.

Using the same one-sided 95\% convention, the complementary upper bound on the failure probability $q=1-p$ is
\[
q_U = 1 - p_L = 1 - 0.05^{1/604} \approx 0.00495,
\]
or approximately 0.495\%. A large-$n$ approximation gives $q_U \approx -\ln(0.05)/604 \approx 0.00496$, which is consistent with the exact calculation.

In practical terms, this means that even though we observed zero failures in 604 cases, statistical uncertainty remains finite. Based on the binomial confidence analysis, we can state with 95\% confidence that the true failure rate for prompt--model cases drawn from this evaluated four-backend mixture is below approximately 0.5\%, and equivalently that the true convergence probability exceeds approximately 99.5\%. While this does not imply perfect reliability, it indicates that production-conformance convergence is highly probable for SysMBench-style prompts under the evaluated setup.

We emphasize that these bounds apply to SysMBench-style prompt distributions under the evaluated controller, conformance checker, and backend configuration. They do not imply universal convergence over arbitrary natural-language inputs. At the same time, the absence of observed failures is consistent with the structure of the loop. Accepted cases terminate immediately, and rejected cases receive deterministic diagnostics that guide subsequent revisions. Within this benchmark distribution, conformance-gated refinement exhibits stable and highly reliable convergence behavior.

\FloatBarrier

\subsection{Grammar Validity vs. Production Conformance}

In practice, a SysMLv2 artifact must conform in a modeling environment before it can be used downstream. Grammar parsability alone is not sufficient. To quantify the gap between these two notions of conformance, we compared ANTLR parsing outcomes against production-conformance outcomes for the initial candidate ($k=0$) across all 604 prompt--model cases.

As shown in Figure~\ref{fig:antlr_vs_production_flow}, of the 604 initial candidates, 369 (61.09\%) passed grammar-level parsing, whereas only 309 (51.16\%) passed production conformance. In 60 cases, the generated model passed ANTLR parsing but failed production conformance. These 60 cases represent 16.26\% of grammar-valid artifacts (60/369) and 9.93\% of all initial outputs (60/604). In other words, grammar-level parsing overestimates operational usability by approximately 16\% among parseable artifacts. We did not observe any case in which a model passed production conformance while failing grammar parsing (0/604), indicating that grammar validity is necessary but not sufficient for production acceptance under the evaluated configuration.

\begin{figure}[H]
\centering
\begin{tikzpicture}[x=1.25cm,y=1.25cm]
\draw[line width=0.65pt] (0,0) rectangle (2,2);
\draw[line width=0.65pt] (1,0) -- (1,2);
\draw[line width=0.65pt] (0,1) -- (2,1);

\node[font=\small\bfseries] at (1.0,2.50) {SysIDE};
\node[font=\footnotesize] at (0.5,2.24) {Fail};
\node[font=\footnotesize] at (1.5,2.24) {Pass};

\node[font=\small\bfseries, rotate=90] at (-0.78,1.0) {ANTLR};
\node[font=\footnotesize, align=right] at (-0.24,1.5) {Fail};
\node[font=\footnotesize, align=right] at (-0.24,0.5) {Pass};

\node[font=\Large] at (0.5,1.5) {235};
\node[font=\Large] at (1.5,1.5) {0};
\node[font=\Large] at (0.5,0.5) {\textbf{60}};
\node[font=\Large] at (1.5,0.5) {309};
\end{tikzpicture}
\caption{Initial-candidate grammar (ANTLR) vs. production conformance (SysIDE), $n=604$. The bolded operational-gap cell (ANTLR pass, SysIDE fail) contains 60 cases.}
\label{fig:antlr_vs_production_flow}
\end{figure}

\FloatBarrier

\subsection{Insights from Extended Analyses}
First, we observe no meaningful relationship between iterations-to-success and either SysMBench difficulty or the length of the generated SysML output. Repair effort therefore does not appear to scale with model size or benchmark difficulty. One possible explanation is that raw line count does not capture the number of distinct grammatical structures that must be correctly formed; shorter outputs may still contain complex reference or typing relationships that trigger conformance-checker failures.

Second, when conformance-checker errors persist across repair iterations, the majority occur because the LLM does not modify the conformance-checker-highlighted code region at all, effectively ignoring the reported diagnostic. In the remaining cases, the model appears to attempt a repair but produces another invalid construction. This pattern suggests both a pipeline efficiency gap, in which repair iterations are wasted when the highlighted region is not revised, and a model limitation, in which the model attempts a repair but fails to generate a syntactically valid correction.

\subsection{Open-Source Dataset Release}

We release the full trajectory-level corpus generated in this study rather than only baseline and final artifacts. The dataset spans the complete SysMBench prompt set of 151 prompts evaluated across four model backends, yielding 604 prompt--model cases. For each case, we retain every conformance-guided refinement step. In total, the release contains 1,043 stored iteration artifacts. Each artifact includes the generated SysMLv2 text and the corresponding production-conformance diagnostics. As a result, the dataset preserves the full correction trajectory for each case, not just its endpoints.

At the artifact level, we classify iterations as positive if they pass production conformance and negative if they fail. Under this definition, the release contains 604 positive artifacts and 439 negative artifacts. For a newly standardized language such as SysMLv2, where publicly available corpora of production-conforming models remain limited, these 604 positive artifacts provide a meaningful resource. The 439 negative artifacts correspond to intermediate repair states within the same 604 trajectories; they do not represent additional independent benchmark instances, but rather the correction path taken for each prompt--model pair.

Releasing full stepwise traces enables analyses that are not possible with endpoint-only datasets. Researchers can study repair dynamics under deterministic conformance-checker feedback, model convergence behavior across trajectories, and explore learning strategies that operate on intermediate states rather than final outputs alone. Because we retain runtime and token-level metadata at both iteration and trajectory levels, the dataset also supports systematic cost--reliability tradeoff analyses under a consistent production-conformance criterion.

To support reproducibility in addition to dataset reuse, the repository release also includes the code, run artifacts, and analysis outputs used in this study, together with the scripts required to regenerate the reported campaign statistics, tables, and figures from the released trajectory data~\cite{LaVoieSupportingRepo}.

\section{Discussion}

\textbf{Production Conformance:} By placing a production conformance checker inside the generation loop, we address a practical barrier to using natural-language--to--SysMLv2 generation in real MBSE workflows. Rather than producing text that appears structurally plausible, we require that every generated model satisfy the same production checks needed to load and use it in an industrial modeling environment. Across 604 prompt--model cases, conformance-guided refinement increases production-conformance from 51.16\% in single-shot generation to 100.00\%. In doing so, we remove the primary syntactic obstacle to integrating LLM-assisted modeling into existing toolchains.

\textbf{Reliability Mechanism:} Shifting from single-shot generation to conformance-guided refinement also changes where reliability is enforced. In the single-shot setting, acceptance depends on whether the language model happens to produce a structurally correct artifact on its first attempt. In the conformance-guided setting, we instead rely on a deterministic backend that evaluates and reports concrete modeling errors. At each iteration, we check the candidate model, revise it based on explicit diagnostics, and re-evaluate it until no conformance errors remain. Because we apply the same production conformance across all runs, acceptance becomes largely independent of backend-specific first-pass behavior. Model providers may differ in cost and latency, but we enforce production acceptance through a shared conformance layer.

\textbf{Deployment Implications:} From a deployment perspective, we believe this distinction is critical. In typical MBSE environments, engineers must load, render, and validate models before using them for visualization, simulation, or verification. Without conformance guidance, generated models are usable only when they happen to meet these constraints. With conformance guidance, we enforce usability as part of the generation process itself. This allows us to incorporate LLM-assisted modeling into day-to-day engineering practice while maintaining bounded operational risk. In this setting, engineers can spend less time correcting structural syntax errors and more time reasoning about architecture, requirements, and verification intent.

\textbf{Convergence Dynamics:} The observed convergence behavior further supports this interpretation. Deterministic diagnostics provide stable correction signals, and we find that most recoverable failures resolve within a small number of repair cycles ($T_{90}=2$, $T_{95}=3$, $T_{99}=4$). Initial candidates often contain multiple structural errors, yet the loop typically resolves these issues in only a few iterations. These results suggest that the refinement process corrects shallow but compound structural inconsistencies rather than exhibiting unstable or oscillatory dynamics.

\textbf{Scope Limits:} We emphasize that the scope of these results is intentionally limited. By enforcing production conformance, we guarantee structural and operational compatibility with a modeling tool, but we do not guarantee that the resulting model faithfully represents the intended system or satisfies its requirements. We do not evaluate semantic correctness, behavioral adequacy, or trace completeness in this study. In addition, we bound our empirical findings to the SysMBench prompt distribution and to a single production conformance backend. Related to our reliance on the SysMBench dataset, we reiterate that our contribution is on production-conformant generation of structural model content given natural language descriptions, rather than guaranteed synthesis of all SysMLv2 diagram types. The accepted outputs in SysMBench overwhelmingly contain descriptions of packages, part definitions, part usages, ports, and connections; corresponding most closely to package, block definition, and internal block definition diagrams in SysML context. A very small number of data points included requirement and parametric syntax. Hence, we do not claim that the generated artifacts are equivalent to sufficiently complete architectural models that retain full decomposition information nor explicit function to form allocation. Future work will examine how the proposed approach could be extended for automating systems architecture work.

Within these bounds, we view conformance-driven refinement as a foundational reliability layer for AI-assisted MBSE. Once we guarantee structural acceptance, we can introduce additional mechanisms such as semantic validation, cross-view consistency analysis, simulation-based checks, and requirement trace evaluation. By establishing production conformance acceptance as a stable baseline, we create a controlled starting point for building higher-level assurances on top of syntactic convergence.
\section{Conclusion}

Our results indicate that single-shot LLM generation does not reliably produce SysMLv2 models that pass production conformance, and this limits the direct use of natural-language--to--SysMLv2 pipelines in real MBSE workflows. In this work, we address this limitation by placing the production conformance inside the generation loop. We generate a candidate model, run it through a production SysMLv2 conformance-checker, and revise it based on the reported diagnostics until the conformance-checker reports zero errors. In doing so, we treat production acceptance as the stopping condition of the generate--check--repair process. Unlike grammar-level approaches that rely on parser checks alone, our method requires acceptance by an industrial modeling tool.

Across the full SysMBench prompt set of 151 prompts and four model backends, yielding 604 prompt--model cases, this approach increases production-conformance acceptance from 51.16\% in single-shot generation to 100.00\% after conformance-guided refinement. Most failures resolve within a small number of repair cycles.

From a practical standpoint, this work provides a reliability layer between LLM-generated text and MBSE tools. Instead of relying on a single pass of generation, we use deterministic conformance-checker feedback to drive correction until the model can be loaded and used in a production environment. This approach does not require retraining model weights and remains agnostic to the underlying language model. To support further research, we release the full trajectory-level data from the campaign, comprising 1,043 iteration artifacts with associated conformance traces and run metadata~\cite{LaVoieSupportingRepo}.

Our results are intentionally scoped to syntactic and operational acceptance under a production conformance-checker. Passing conformance is necessary for deployment, but it does not guarantee semantic correctness or architectural adequacy. Future work should examine cross-conformance replication, semantic alignment with requirements, and task-level evaluation of behavioral correctness. We view production-conformance enforcement as a first step toward reliable NL-to-MBSE integration, not the final step.
\section{Limitations and Future Work}

While the proposed conformance-checker-in-the-loop framework achieves
production-conformance acceptance across all benchmark cases,
several limitations remain.

First, our study focuses exclusively on syntactic and production-level
conformance. Passing a production conformance-checker ensures that a model can
be loaded, rendered, and processed by downstream tools, but it does
not guarantee semantic correctness, architectural adequacy, or
requirements fidelity. A model may satisfy all name resolution,
typing, and ownership constraints while still misrepresenting the
intended system behavior. Extending the framework to incorporate
requirement-level checks, trace consistency, and behavior validation
is an important next step.

Second, our experiments rely on a single production conformance-checker,
SysIDE. Although SysIDE reflects industrial conformance behavior,
different toolchains may enforce additional constraints or differ in
diagnostic reporting. Cross-tooling replication would strengthen
the generality of the approach and help characterize how tooling
differences affect repair trajectories and convergence behavior.

Third, the empirical results are scoped to the SysMBench prompt
distribution. While the benchmark spans 151 prompts and multiple
model backends, it does not capture the full variability of industrial
natural-language specifications, including noisy, incomplete, or
internally inconsistent requirements. Evaluating robustness under
longer and less structured inputs would provide further insight into
deployment readiness.

Fourth, although convergence is observed empirically in all 604
prompt--model cases, we do not provide a formal guarantee of
convergence. The observed contraction-like behavior in early repair
cycles suggests stable dynamics under deterministic diagnostics,
but a theoretical analysis of convergence conditions and potential
failure modes remains open.

Finally, we do not quantify cost--reliability tradeoffs in detail.
Conformance-guided refinement introduces additional model calls and
runtime overhead. A systematic study of iteration counts, token
usage, and latency under varying prompt complexity would clarify
the practical operating envelope of the approach.

Future work will focus on three directions. First, we will extend the
conformance-checker-in-the-loop paradigm to incorporate semantic and
simulation-based checks, moving from syntactic acceptance to
task-level correctness. Second, we will investigate cross-conformance
evaluation to assess toolchain sensitivity and strengthen claims of
deployment robustness. Third, we will explore adaptive repair
strategies that prioritize high-impact diagnostics to reduce iteration
count and cost. Together, these efforts aim to build a layered
assurance framework for reliable natural-language--to--MBSE
generation.

\bibliographystyle{unsrtnat}
\bibliography{references_submission}

\appendix
\section{Auxiliary Demonstration: Grammar Parsability vs. Production Conformance}
\label{appendix:antlr_vs_syside}

To support the design choice of using production conformance, rather than grammar parsing alone, as the acceptance oracle, we ran an auxiliary demonstration included in the repository~\cite{LaVoieSupportingRepo}. This demonstration is not a second primary experiment; it shows that parser acceptance and production conformance are distinct outcomes.

We evaluated 10 intentionally distinct SysMLv2 examples located in the Appendix~A demonstration directory of the released repository, under \texttt{examples/mismatch\_10\_distinct/}. Each file was designed to remain grammar-parseable while violating a model-wide constraint typically enforced by a production conformance checker. The evaluation pipeline was:
\begin{enumerate}
    \item ANTLR parse check using the SysMLv2 ANTLR4 parser (generated grammar derived from the SysML v2 Pilot Implementation)~\cite{hamrSysmlParser2026,sysmlV2PilotImplementation2026}.
    \item Production conformance check using SysIDE.
\end{enumerate}

Results were unambiguous: all 10/10 examples passed ANTLR parsing, while 0/10 were accepted by production conformance (10/10 mismatch cases). Conformance-checker diagnostics were dominated by unresolved-reference failures (9/10, \texttt{reference-error}), with one invocation-typing failure (1/10, \texttt{invocation-expression-instantiated-type}). This pattern, summarized in Table~\ref{tab:antlr_vs_syside_10cases}, directly illustrates that context-free syntax conformance is necessary but not sufficient for operational model acceptance in production modeling environments.

\subsection*{Concrete Example}
The following example is grammar-parseable but fails production conformance:
\begin{quote}
\small\ttfamily
package Demo07 \{\\
\ \ part def Wheel \{\\
\ \ \ \ attribute hubDiameter: LengthValue;\\
\ \ \ \ part tire \{ attribute width: LengthValue; \}\\
\ \ \ \ attribute outerDiameter: LengthValue = (hubDiameter + 2 * tire.height);\\
\ \ \}\\
\}
\end{quote}

ANTLR parsing accepts the structure. Production conformance rejects it with:
\begin{quote}
\small\ttfamily
error (reference-error): No Feature named 'height' found.
\end{quote}

\begin{table}[tbp]
\centering
\caption{Ten-case auxiliary demonstration: all examples parse under ANTLR, none pass production conformance.}
\label{tab:antlr_vs_syside_10cases}
\scriptsize
\setlength{\tabcolsep}{2pt}
\begin{tabular}{p{0.08\columnwidth} p{0.49\columnwidth} p{0.12\columnwidth} p{0.21\columnwidth}}
\toprule
\textbf{ID} & \textbf{Injected Condition} & \textbf{ANTLR} & \textbf{SysIDE} \\
\midrule
01 & Missing imported namespace & Pass & Fail (reference) \\
02 & Missing port type & Pass & Fail (reference) \\
03 & Missing attribute type & Pass & Fail (reference) \\
04 & Missing specialization base & Pass & Fail (reference) \\
05 & Action typed by undefined behavior type & Pass & Fail (reference) \\
06 & Invocation target is not a behavior & Pass & Fail (invocation) \\
07 & Missing referenced feature in expression & Pass & Fail (reference) \\
08 & Missing root-qualified namespace & Pass & Fail (reference) \\
09 & Redefinition of missing feature & Pass & Fail (reference) \\
10 & Missing namespace in type use & Pass & Fail (reference) \\
\bottomrule
\end{tabular}
\end{table}

\lstdefinestyle{appendixstderr}{
    basicstyle=\ttfamily\scriptsize,
    breaklines=true,
    columns=fullflexible,
    keepspaces=true,
    showstringspaces=false,
    frame=single,
    framerule=0.4pt,
    framesep=4pt,
    xleftmargin=0pt,
    xrightmargin=0pt,
    aboveskip=0.4em,
    belowskip=0.4em
}

\section{Extended Analysis: Difficulty and Output-Length Signals}
This appendix section examines whether iteration count is meaningfully related to either the SysMBench difficulty label or the length of the generated SysML output. The purpose is to test whether repair effort is primarily a size-driven phenomenon.

\subsection{SysMBench Difficulty Versus Iterations-to-Success}
\begin{figure}[H]
\centering
\captionsetup{width=\textwidth}
\includegraphics[width=\textwidth,keepaspectratio]{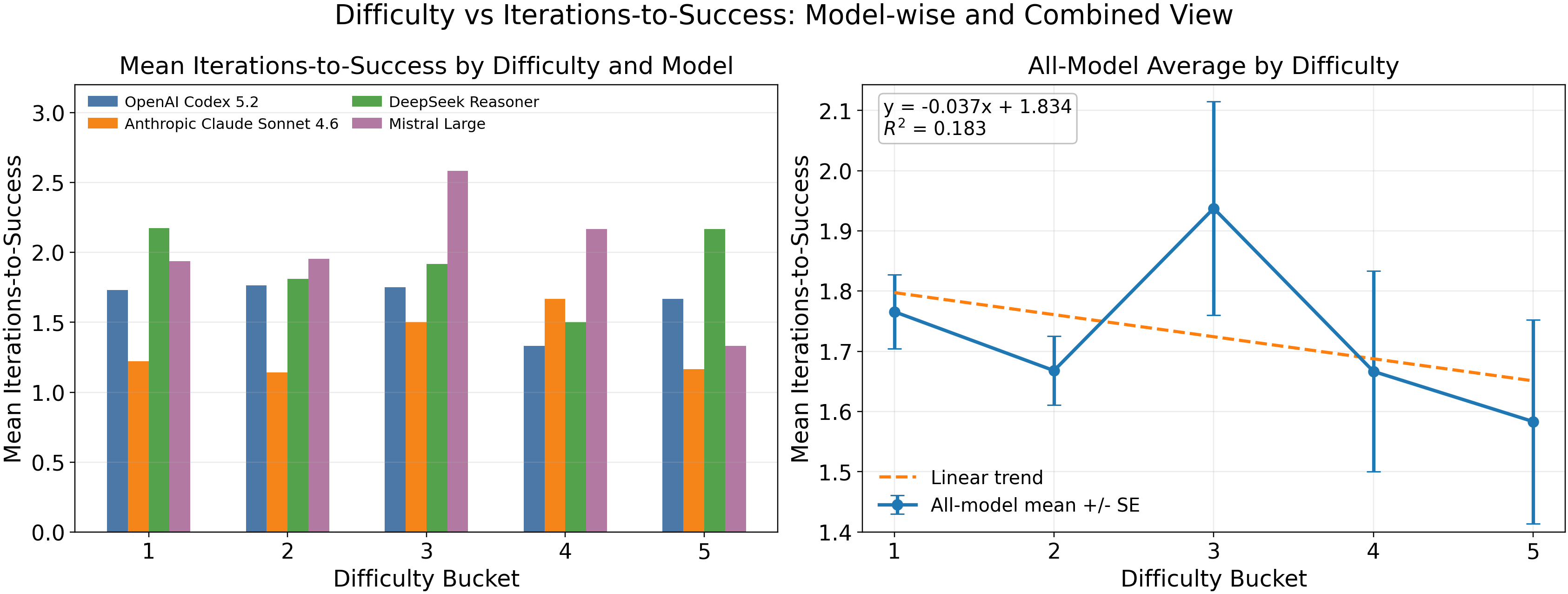}
\caption{Difficulty-conditioned repair effort with model split and pooled trend. The pooled mean iterations-to-success are 1.766, 1.668, 1.938, 1.667, and 1.583 for difficulty levels 1 through 5, respectively, with a pooled linear fit of \(R^2 \approx 0.183\).}
\label{fig:appendix_difficulty}
\end{figure}

In this first analysis, we use the SysMBench difficulty classification~\cite{jin2025sysmbench}. In their scheme, each prompt is assigned a difficulty bucket from the line count of the hand-authored ground truth SysML model, not from the generated output. The bucket definitions are: difficulty 1 for fewer than 30 lines, difficulty 2 for 30--59 lines, difficulty 3 for 60--89 lines, difficulty 4 for 90--119 lines, and difficulty 5 for 120 or more lines. Across the 151 benchmark prompts, this yields \(n=63, 64, 12, 6,\) and \(6\) prompts in difficulty buckets 1 through 5, respectively. Because each prompt is evaluated with four models, the pooled right-panel averages in Figure~\ref{fig:appendix_difficulty} are computed from 252, 256, 48, 24, and 24 prompt--model cases across the five difficulty levels. Figure~\ref{fig:appendix_difficulty} tests whether that benchmark-defined size proxy tracks repair effort in the conformance-checker-in-the-loop setting.

The pooled mean iterations-to-success are 1.766, 1.668, 1.938, 1.667, and 1.583 across difficulty levels 1 through 5, respectively. The weak linear fit indicates that the hand-authored ground truth SysML line-count difficulty label is not strongly correlated with repair effort, as measured here by iterations-to-success.

\FloatBarrier

\subsection{Generated Output Length Versus Iterations-to-Converge}
Since SysMBench difficulty is line-count based, and one might reasonably expect repair effort to increase with the number of lines generated, it is useful to ask whether the length of the \emph{generated} SysML output is itself correlated with iterations-to-converge.

\begin{figure}[H]
\centering
\captionsetup{width=\textwidth}
\includegraphics[width=\textwidth,keepaspectratio]{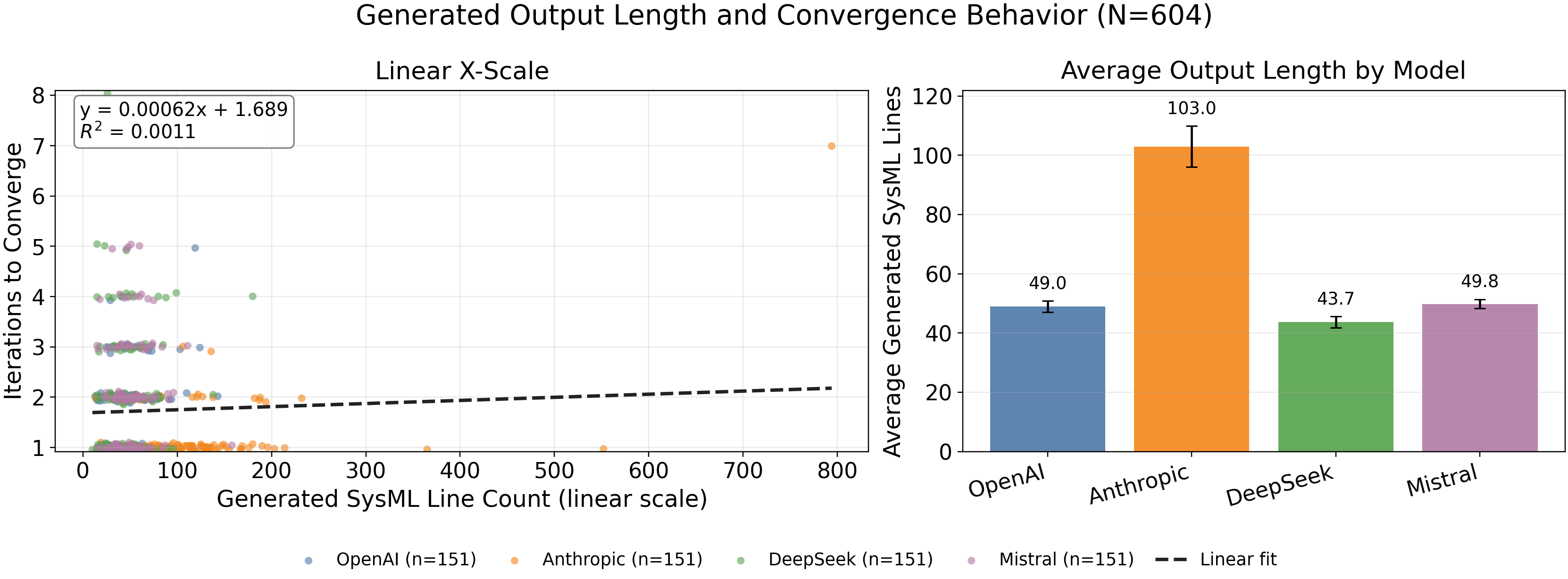}
\caption{Generated SysML output length analysis. The left panel shows a scatter plot of generated SysML line count versus iterations-to-converge, with the pooled linear fit overlaid (\(R^2 = 0.0011\), slope \(= 0.00062\)). The right panel shows the average generated SysML line count for each model, with error bars indicating standard error.}
\label{fig:appendix_generated_lines}
\end{figure}

The pooled linear fit in Figure~\ref{fig:appendix_generated_lines} is essentially flat (\(R^2 = 0.0011\), slope \(= 0.00062\)), which indicates that generated SysML line count explains almost none of the variation in iterations-to-converge. In this analysis, generated output length is therefore not strongly correlated with repair effort.

The model-level averages in the right panel reinforce the same point. OpenAI, DeepSeek, and Mistral produce broadly similar output lengths on average (approximately 49, 44, and 50 lines, respectively), whereas Anthropic produces about 103 lines on average, or roughly twice as many. However, Anthropic also required the fewest iterations-to-success on average in the main results (1.23, versus 1.73 for OpenAI, 1.97 for DeepSeek, and 1.98 for Mistral), further indicating that line count alone is a poor standalone proxy for repair effort. Taken together, Figures~\ref{fig:appendix_difficulty} and~\ref{fig:appendix_generated_lines} suggest that repair effort is not well explained by size alone.

One possible explanation is that raw line count does not necessarily reflect the number of distinct syntactic structures that must be formed correctly. A model can generate many lines by expanding a relatively uniform pattern, whereas a shorter output may still involve a wider variety of declarations, references, and nested relationships that are more sensitive to conformance failure. Simply producing more lines of the same structure therefore need not imply greater repair effort.

\FloatBarrier

\section{Extended Analysis: Classifying Repair Attempts on Persistent Errors}
This appendix section looks at what the LLM does from one iteration to the next when a persistent error is still present. We treat an error as \emph{persistent} if an identical SysIDE diagnostic, meaning the same error family and the same \texttt{stderr} message, appears in two consecutive iterations of the same prompt--model run. To study the repair attempt itself, we compare the SysML artifact from iteration \(k\), which is the previous attempt fed back into the loop, with the revised SysML artifact produced by the LLM at iteration \(k+1\). This analysis uses the same 604 SysMBench prompt--model trajectories reported in the main results.

\subsection{Error Transition Outcome Explanations}
Using the SysIDE \texttt{stderr} outputs, we ask a simple question: when an identical error is still present at iteration \(k+1\), did the LLM appear to change the highlighted failing region, or did it leave that region unchanged? Across all back-to-back iteration pairs, this yields 1502 exact-error transitions. Each transition corresponds to one exact SysIDE diagnostic observed at iteration \(k\) and checked again at iteration \(k+1\). This is smaller than the total number of raw error instances because repeated copies of an identical SysIDE message within a single iteration are collapsed into one signature-level transition event. That choice is intentional: here we want to measure whether the model corrected an error pattern from one iteration to the next, not whether it repaired every repeated occurrence independently. We interpret many repeated identical grammar diagnostics as manifestations of the same underlying structural issue, so once the model applies the relevant pattern-level fix, repeated instances will often be corrected together rather than one by one.

Of these 1502 transitions, 1262 are resolved by the next iteration and are therefore not persistent. The remaining 240 persist through the next repair attempt. These 240 persistent transitions are the cases for which we want to understand whether the model tried to repair the highlighted region but failed, or whether it effectively carried the same failing code forward unchanged.

We then classify the 240 persistent cases into one of two categories:
\begin{itemize}
    \item \textbf{Unaddressed and not fixed}: an identical conformance error remains at iteration \(k+1\), and the conformance-checker highlights an identical code snippet again.
    \item \textbf{Addressed but not fixed}: an identical conformance error remains at iteration \(k+1\), but the conformance-checker now highlights a different code snippet instead, suggesting that the LLM changed the code to attempt to fix the error, but was unsuccessful.
\end{itemize}

To classify these persistent cases, we use the \texttt{stderr} outputs from SysIDE directly. For each persistent identical conformance error, we compare the SysIDE \texttt{stderr} block at iteration \(k\) with the corresponding \texttt{stderr} block at iteration \(k+1\), and check whether SysIDE is still highlighting the identical code snippet or now highlights a different snippet instead.

\noindent\begin{minipage}{\columnwidth}
For example, consider the following SysIDE \texttt{stderr} output. For readability, the long expected-token list is abbreviated here, but the full raw message is used in the analysis:
\begin{lstlisting}[style=appendixstderr]
iteration_01.sysml:36:19: error (parsing-error):
    Unexpected 'state', expected one of [...]
    36 |         attribute state : VehicleState;
       |                   ^^^^^
\end{lstlisting}
In this example, we classify the transition as ``unaddressed and not fixed'' only if iteration \(k+1\) still contains that identical \texttt{parsing-error} message and still highlights the identical code snippet, \texttt{attribute state : VehicleState;}. If iteration \(k+1\) still contains the identical \texttt{parsing-error} message but SysIDE now highlights a different code snippet instead, we classify the transition as ``addressed but not fixed.'' This is determined directly from conformance stderr rather than by manually reading the full file.
\end{minipage}

A small amount of noise is possible for the ``addressed but not fixed'' category if multiple identical errors occur in different code regions within the same file. In that edge case, the LLM could fix one identical occurrence but fail to fix another identical occurrence on a different code segment, which would still appear here as ``addressed but not fixed.'' We expect this to be uncommon in practice, because many grammar repairs are pattern-level corrections that tend to fix repeated identical structures together.

\subsection{Persistent Error Analysis}
\begin{figure}[H]
\centering
\captionsetup{width=0.86\textwidth,skip=2pt}
\begin{tikzpicture}
\begin{axis}[
    xbar,
    width=0.66\columnwidth,
    height=0.18\columnwidth,
    xmin=0,
    xmax=105,
    xlabel={Share of Persistent Transitions (\%)},
    symbolic y coords={Addressed,Unaddressed},
    ytick={Unaddressed,Addressed},
    yticklabels={{Unaddressed\\and not fixed},{Addressed\\but not fixed}},
    enlarge y limits={abs=4pt},
    axis lines*=left,
    xmajorgrids=true,
    grid style={dashed,gray!25},
    tick label style={font=\footnotesize},
    yticklabel style={align=right},
    label style={font=\footnotesize},
    nodes near coords={\pgfmathprintnumber[fixed,precision=2]{\pgfplotspointmeta}},
    every node near coord/.append style={font=\footnotesize, text=black, anchor=west, xshift=2pt},
]
\addplot+[fill=singleShotColor, draw=singleShotColor!70!black] coordinates {
    (64.58,Unaddressed)
};
\addplot+[fill=pipelineColor, draw=pipelineColor!70!black] coordinates {
    (35.42,Addressed)
};
\end{axis}
\end{tikzpicture}
\caption{Persistent exact-error outcomes for the 240 transitions where the same conformance error remains at iteration \(k+1\): 155/240 unaddressed and not fixed; 85/240 addressed but not fixed.}
\label{fig:appendix_persistent_transition_outcomes}
\end{figure}
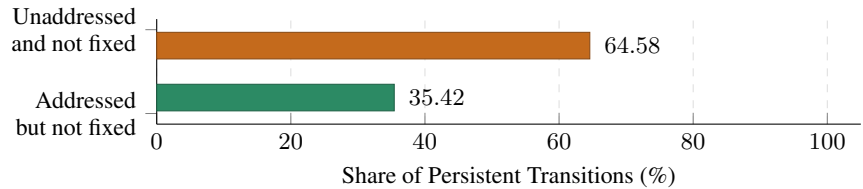

\enlargethispage{2\baselineskip}
Figure~\ref{fig:appendix_persistent_transition_outcomes} restricts attention to the 240 persistent transitions and shows the split between the two persistent outcomes. Of these persistent cases, 155/240 (64.6\%) are ``unaddressed and not fixed,'' while 85/240 (35.4\%) are ``addressed but not fixed.''

The larger unaddressed-and-not-fixed share points to an inefficiency in the repair loop. In these cases, SysIDE is still highlighting the identical offending code region in the next iteration, suggesting that the LLM did not change that highlighted region after receiving the conformance-checker feedback. This motivates future work on stronger grounding of the prompt to the conformance-checker-highlighted snippet, or repair policies that explicitly require the highlighted region to be revised before resubmission.

The addressed-but-not-fixed share instead suggests LLM limitations. In these cases, the model appears to make a repair attempt, but the identical conformance error still remains. This suggests a limitation in repair capability rather than simple loop inefficiency: the model changes the code, but does not produce a syntactically valid correction. Future work here should focus on constrained editing or fine-tuning with the large datasets of syntactically valid code that can now be produced. One caveat is that this analysis is based on SysIDE \texttt{stderr} and highlighted snippets, so rare repeated-identical-error cases can still add a small amount of noise.

\end{document}